# Labeling schemes for tetrahedron equations and dualities between them

Jarmo Hietarinta[*]
L.P.T.H.E.[†], Tour 16, 1$^{\text{er}}$ étage, 4, place Jussieu
F-75252 Paris Cedex 5, France

February 28, 1994

## Abstract

Zamolodchikov's tetrahedron equations, which were derived by considering the scattering of straight strings, can be written in three different labeling schemes: one can use as labels the states of the vacua between the strings, the states of the string segments, or the states of the particles at the intersections of the strings. We give a detailed derivation of the three corresponding tetrahedron equations and show also how the Frenkel-Moore equations fits in as a *nonlocal* string labeling. We discuss then how an analog of the Wu-Kadanoff duality can be defined between each pair of the above three labeling schemes. It turns out that there are two cases, for which one can simultaneously construct a duality between *all* three pairs of labelings.

---

[*]On leave of absence from Department of Physics, University of Turku, FIN-20500 Turku, Finland
Email: hietarin@utu.fi
[†]Laboratoire Associé au CRNS UA 280

# 1 Introduction

Now that a quite good understanding of 1+1 dimensional integrable systems (both classical and quantum, continuum and discrete) has been obtained, the attention has turned to higher dimensions where serious difficulties have been encountered. The various approaches that were successful in 1+1 dimensions have different natural extensions to 2+1 dimensions and it is not clear which method is the best. It is therefore important to push each one and hopefully they will illuminate different aspects of the 2+1 integrable systems.

In this paper we consider the extension of the Yang-Baxter/star-triangle equations to 2+1 dimensions. The fundamental work in this problem was done already at the beginning of the 1980's, first by Zamolodchikov [1], who derived the relevant tetrahedron equations by studying the scattering of straight strings. In this formulation it was natural to use the quantum numbers of the string segments (faces in the lattice formulation) as labels. Subsequently Bazhanov and Stroganov [2] wrote down the equations corresponding to cell and edge labeling. A different type of edge tetrahedron equation has been proposed by Frenkel and Moore [3].

Higher dimensional generalizations of the tetrahedron equations have also been discussed in the literature. The 4-simplex equations appeared already in the above paper of Bazhanov and Stroganov [2] and d-simplex equations have been discussed e.g. by Maillet and Nijhoff [4] and Carter and Saito [5]. For related geometric constructions, see [6].

In all formulations the number of equations is huge even for the simplest two-state model, and subsequent progress has been slow because it has been exceedingly difficult to find solutions (especially those with a spectral parameter) to these equations. The original solution proposed by Zamolodchikov [1] has been studied further by Baxter [7], and only quite recently some further solutions have been found [8].

Much of the work on tetrahedron equations has been done in the framework of solvable lattice models. Each formulation has its own natural properties and from time to time it is useful to look at all of them for inspiration. With this in mind we return to the original formulation of straight string scattering, and from this point of view look in detail on the properties of the various methods of labeling the tetrahedron equations.

Our main objective is to study the analogues of Wu-Kadanoff duality in the tetrahedron situation. The duality between the star-triangle and Yang-Baxter equations imposes certain restriction for the existence of nonzero elements of the $R$-matrix. These restrictions amount to the very important 8-vertex ansatz, and our hope is that also in the tetrahedron case duality will lead us to fruitful ansatze.

The organization of this paper is as follows: In the next section we start by rederiving the Yang-Baxter and star-triangle equations in detail, because we want to use the analogies in the tetrahedron case. We discuss also in similar detail the well know Wu-Kadanoff duality that connects the Yang-Baxter and star-triangle equations under certain circumstances. In Sec. 4 we derive then three versions of the tetrahedron equations. They differ by the choice of labels, we can use as labels the state of the vacuum between the strings, the state of the string segments, or the state of the particles at the intersection of the strings. (We will ignore the spectral parameters, because they are not relevant to the labeling problem.) The Frenkel-Moore equation will also be obtained if we use nonlocal string labeling. In Sec. 5 we will derive a duality between each pair of tetrahedron equations, in analogue to the Wu-Kadanoff duality. This implies certain restrictions (which are explicitly written out) on the functions. When these restrictions hold both equations reduce to a common equation that has fewer labels and summation indices. Finally we consider the possibility of one equation being dual to the two others.

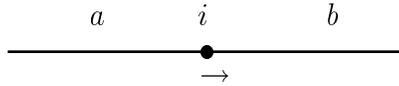

Figure 1: The basic state and its labeling.

Our main result is the indentification of two cases, for which there can simultaneously exist a duality between each pair of labeling schemes.

## 2  The Yang-Baxter and star-triangle equations

In this section we review the particle scattering derivation of Yang-Baxter and star-triangle equations. This is done in some detail, because analoguous steps will be followed in the derivation of the tetrahedron equations.

### 2.1  The basic state and the basic scattering process

The Yang-Baxter equation can be interpreted as describing scattering, with a factorizable scattering matrix, in 1+1 dimensions [9]. The ambient space is 1-dimensional and the fundamental state is a point particle moving with a constant speed, see Fig. 1. The particle divides the space into two parts and the state of the vacuum can be different in them (in this case the moving particle can be modeled by a kink that interpolates between the two vacua). In the general case we must therefore use three labels (in addition to a dynamical characterization by velocity) to completely describe the basic state.

The basic scattering process is obtained if we have two particles moving with different velocities. Initially the particles approach each other, scatter, and finally recede from each other, see Fig. 2. During scattering the momenta do not change, but the internal states of the particles can change, as well as the vacuum between them. Thus in principle the scattering amplitude could at the same time depend on four vacuum labels, four particle labels, and the relative momenta (or the angle between the intersecting trajectories). Often one uses only vacuum or particle labeling, vacuum (or 'face') labeling e.g. in the face formulation of lattice models, where the scattering amplitude of Fig. 2 is given by the 'Boltzmann weight' $w(b, c, d, a; u)$; here the first label is for the vacuum between the incoming particles and thereafter counterclockwise. In particle labeling (or vertex formulation of lattice models) one uses the $R$-matrix and the amplitude of the above process is $R_{ij}^{kl}$.

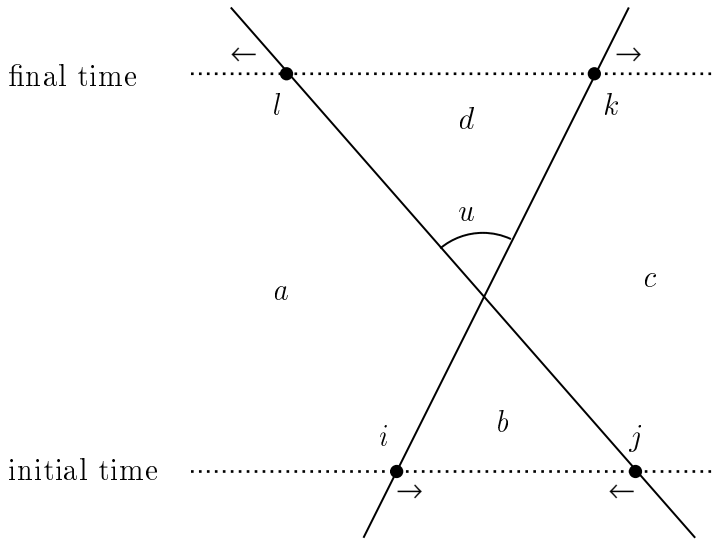

Figure 2: The basic process: Particles $i$ and $j$ collide resulting with particles $k$ and $l$. The collision is elastic in the sense that the momenta do not change and inelastic in that the colliding particles and the vacuum between them can change.

## 2.2 Condition from factorizability

The above describes fully what can happen with two particles. The situation becomes more interesting when we have three particles. Since the two-body scatterings are elastic (momenta do not change), the approaching particles will undergo precisely three of these pairwise scatterings before they fly out again (Fig. 3). The order in which they take place depends on the relative initial location of the particles, and the natural assumption is that for the three particle scattering amplitude *the order should not make any difference*. Pictorially this is stated in Fig. 4 [9].

The equations following from Fig. 4 depend on which labeling scheme we are using. If we use vacuum (face) labeling, where the scattering amplitude of Fig. 2 is given by $w(b, c, d, a; u)$, the condition of Fig. 4 writes

$$\sum_g w(b, c, g, a; u) w(c, d, e, g; u+v) w(g, e, f, a; v) =$$
$$\sum_g w(c, d, g, b; v) w(b, g, f, a; u+v)) w(g, d, e, f; u). \qquad (1)$$

This is the so called 'star-triangle equation' [10]. Note in particular that there is just one summation index, the vacuum inside the triangle, and this index appears in all scattering amplitudes.

If we use particle (vertex) labeling and the scattering amplitude for the scattering process Fig. 2 is given by $R^{kl}_{ij}(u)$ then the condition of Fig. 4 writes

$$R^{k_1 k_2}_{j_1 j_2}(u) R^{l_1 k_3}_{k_1 j_3}(u+v) R^{l_2 l_3}_{k_2 k_3}(v) = R^{k_2 k_3}_{j_2 j_3}(v) R^{k_1 l_3}_{j_1 k_3}(u+v) R^{l_1 l_2}_{k_1 k_2}(u). \qquad (2)$$

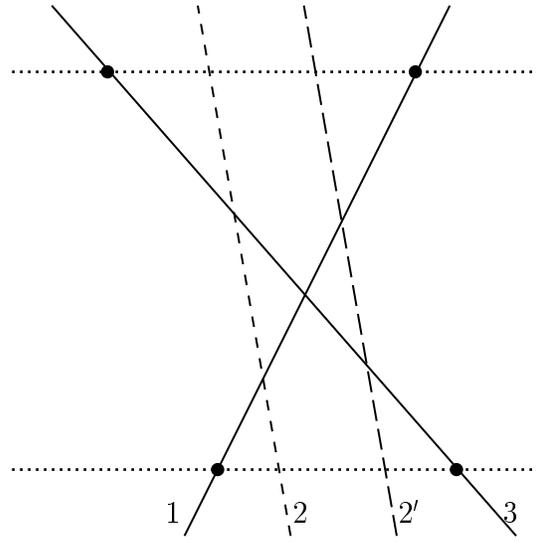

Figure 3: The two possible three particle scattering scenarios with the same momenta but different initial position of particle 2

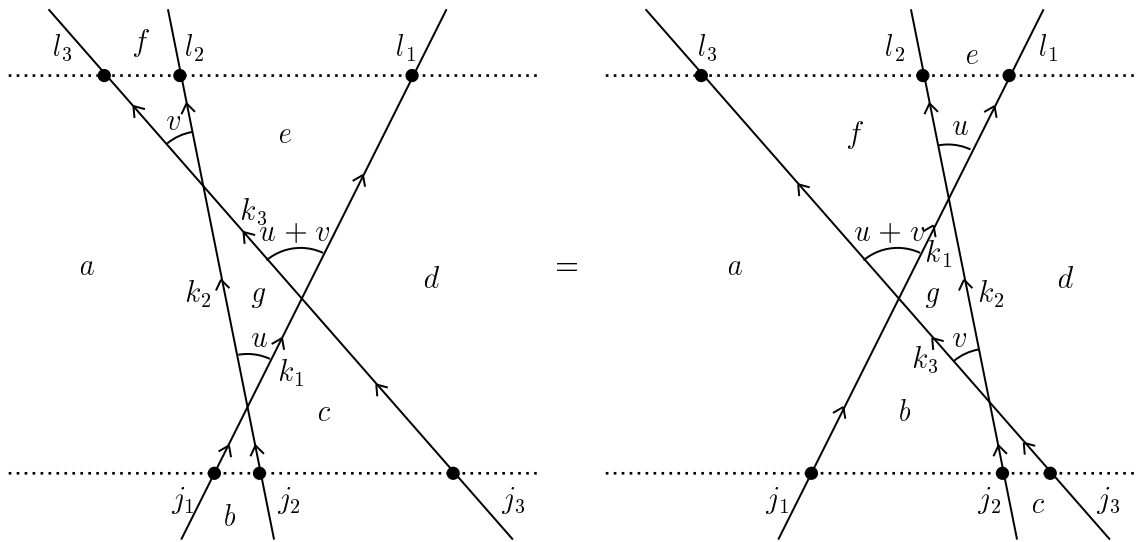

Figure 4: Pictorial representation of the factorization condition.

In this case there are three pairwise summation indices ($k_i$) so we have used Einstein's summation convention. This equation is called the '(quantum) Yang-Baxter equation'.

# 3  The Wu–Kadanoff duality

The vacuum (face) labeling and particle (vertex) labeling are not completely isolated; under certain restrictions (which turn out to be of practical significance) one can establish a well known duality between them [11]. The derivation is again done in detail because the analogies will be used later with the tetrahedron equations.

## 3.1  Generalities

Above the nature of the labels was left open, because no structure was necessary. However, now we have to do some algebra with the indices and thus we have to discuss the domain we will be working in.

Let us first recall the setting of the original lattice work. The face (vacuum) and edge (particle) labels were given by the state of a spin, either up or down. For models with more than two states the values of the state labels could be modeled by integers modulo $d$, or for more exotic models by pairs of integers modulo $d$, etc. In general we assume that the labels are elements of some abelian group $I$. For spins one often uses multiplicative notation for this group, but in this paper we use additive notation.

In the following we have to do some mild arithmetics with the labels. In general the vacuum and particle labels could be from a different abelian group, but to simplify notation we take them to be the same, $I$. We will be needing a linear map from $I$ to $I$ with some parameters, we assume that the parameters belong to ring $\mathcal{R}$ so that $I$ will be an $\mathcal{R}$-module. In fact we will assume that the parameters belong to some (finite) field $\mathcal{F}$. The most common example is that $\mathcal{F}$ = integers modulo a prime number $p$, and $I = \mathcal{F}^n$.

Thus the vacuum labels of the Boltzmann weight $w$ belong to $I_v$ and $w$ itself defines a map $w : I_v^{\otimes 4} \to \mathbb{C}$. Similarly the particle labels of the scattering amplitude belong to $I_p$ and $R : I_p^{\otimes 4} \to \mathbb{C}$.

In order to define a duality relation between two labeling schemes we use the following basic principles:

1. Hierarchy: The scattering amplitude of a labeling scheme with fewer summation indices in the factorization condition is expressed in terms of those whose equation has more summation indices.

2. Locality: The state of a particle is given by its nearest neighbor vacuum states.

3. Linearity: The above dependence is linear.

## 3.2  Details

The above principles imply that the duality between $w$ and $R$ is given by $w = R \circ f$, or explicitly (c.f. Fig. 4)

$$w(b,c,d,a) = R^{\epsilon_3 d + \tau_3 c, \epsilon_4 a + \tau_4 d}_{\epsilon_1 a + \tau_1 b, \epsilon_2 b + \tau_2 c}, \tag{3}$$

with 8 constants $\epsilon_i$, $\tau_i$, that are free at the moment (in the simplest case we expect these constants to be $\pm 1$). This defines the local maps $f_i : I_v^{\otimes 2} \to I_p$, for example $j_1 := f_1(a, b) = \epsilon_1 a + \tau_1 b$, and the function $f : I_v^{\otimes 4} \to I_p^{\otimes 4}$ mentioned above is their extension to the domain of $w$, as given in (3). We must in fact extend $f$ to the full space in question, defining $f^* : I_v^{\otimes 7} \to I_p^{\otimes 9}$, and verify that with this mapping the two equations (1) and (2) reduce to a common equation. To do the above we proceed in four steps:

i) *Label matching:* In (3) there are four different pairs of $\epsilon$ and $\tau$ coefficients, but they cannot all be free, because in (2) the $k_\mu$ indices, for example, appear in different places in $R$ and they must of course have the same expression. For example, from the two $k_1$'s on the LHS we conclude that $\epsilon_3 = \epsilon_1$ and $\tau_3 = \tau_1$. The final result is that the dependence on the subscript is trivial, $\epsilon_i = \epsilon, \tau_i = \tau, \forall i$, i.e. functions $f_i$ are all alike. The extensions to $f$ and $f^*$ follow then straightforwardly.

ii) *Constraints on $R$ and $w$:* In (2) there are three summation indices, in (1) only one; the range of the map $f^*$ cannot, therefore, be the full $I_p^{\otimes 9}$. We can reduce the number of summation indices only by introducing constraints of the type "$R = 0$ for some index values", which means that we must look at the range of $f$ and define $R = 0$ outside. The label restriction on $R$ must in general have the form

$$R_{ij}^{kl} = 0, \text{ unless } Ai + Bj + Ck + Dl = 0,$$

where the constants $A$, $B$, $C$, $D$ are determined by $f$, i.e. the form of $R$ used in (3). The labels $a, b, c, d$ are here free and we need a nontrivial solution of

$$A(\epsilon a + \tau b) + B(\epsilon b + \tau c) + C(\epsilon d + \tau c) + D(\epsilon a + \tau d) = 0,$$

on the whole $I_v^{\otimes 4}$. This is easily found, we get $A = -D = \epsilon$, $B = -C = -\tau$, hence the restriction must be

$$R_{ij}^{kl} = 0, \text{ unless } \epsilon i + \tau k = \epsilon l + \tau j. \tag{4}$$

Since we can determine the upper right label, say, of $R$ from the others we conclude from (3) that $w$ depends only on three indices, a convenient choice is

$$w(b, c, d, a) = \tilde{w}(\epsilon a + \tau b, \epsilon b + \tau c, d - b), \text{ or } R_{ij}^{k*} = \tilde{w}(i, j, \tfrac{1}{\epsilon}(k - j)). \tag{5}$$

Thus $f$ is one-to-one only on the three dimensional subspaces shown in (5)

iii) *Label conversion:* Next we must extend $f$ to $f^*$ and find its kernel and range. For explicit calculations we observe that the constraint (4) is coded into (3) so all its consequences are obtained when we substitute (3) into (1) and try to convert it into (2). This is easy to do: for example, in order to get the first term on the LHS of (2) right we find that we should make the substitutions

$$a = \tfrac{1}{\epsilon} j_1 - \tfrac{\tau}{\epsilon} b, \quad c = \tfrac{1}{\tau} j_2 - \tfrac{\epsilon}{\tau} b, \quad g = \tfrac{1}{\epsilon}(k_1 - j_2) + b.$$

This introduces $j_1, j_2$ and $k_1$ but then $k_2$ is fixed by

$$\epsilon j_1 + \tau k_1 = \epsilon k_2 + \tau j_2. \tag{6}$$

In the second term we can convert $d$ and $e$ into $j_3$ and $l_1$, respectively, but $k_3$ is fixed by

$$\epsilon k_1 + \tau l_1 = \epsilon k_3 + \tau j_3. \tag{7}$$

These are nothing more than (4) applied to the $R$-term in question, but with the third term we get (after converting $f$ into $l_2$) one overall restriction on the free labels

$$\epsilon^2(j_1 - l_3) - \epsilon\tau(j_2 - l_2) + \tau^2(j_3 - l_1) = 0, \tag{8}$$

which decreases the number of equations from $d^6$ to $d^5$. Similar results are obtained on the RHS, the restrictions on the $k$'s are now

$$\epsilon k_3 + \tau j_3 = \epsilon j_2 + \tau k_2, \quad \epsilon l_2 + \tau k_2 = \epsilon k_1 + \tau l_1, \tag{9}$$

while the condition on the free labels is again (8). Thus the range of $f^*$ is $I_p^{\otimes 9}$ subject to (6 -8) on the LHS while on the RHS the restriction is (8,9). A one-to-one mapping is then obtained for a six dimensional subspace, the kernel is generated e.g. by $b$.

iv) *The common equation:* Finally, we can write down the common equation to which (1) and (2) reduce under (5). Using the further redefinitions $k_1 = \epsilon k + j_2$ on the LHS, $k_3 = \tau k + j_2$ on the RHS, and $l_1 = j_3 + \epsilon l_1'$, $l_3 = j_1 + \tau l_3'$ we find

$$\sum_k \tilde{w}(j_1, j_2, k) \, \tilde{w}(j_2 + \epsilon k, j_3, l_1') \, \tilde{w}(j_1 + \tau k, j_2 + \epsilon k + \tau l_1', l_3' - k)$$
$$= \sum_k \tilde{w}(j_2, j_3, k) \, \tilde{w}(j_1, j_2 + \tau k, l_3') \, \tilde{w}(j_2 + \tau k + \epsilon l_3', j_3 + \epsilon k, l_1' - k). \tag{10}$$

In this final equation we have resolved all conditions following from duality. The equation contains only one summation index, the functions depend only on three labels, and all the labels are free (because $l_2$ does not appear). Thus we have $d^5$ equations for $d^3$ unknowns. Furthermore, the reflection symmetry of (2) is preserved: the right and left sides are exchanged if we exchange the first two labels of $\tilde{w}$ and change $1 \leftrightarrow 3$ and $\epsilon \leftrightarrow \tau$.

In the Wu-Kadanoff duality [11] one takes $\epsilon = -\tau = 1$. This means that if two adjacent cell spins point to the same direction the edge spin in between points up. For $R$ this implies that $i + j = k + l$ is the necessary condition for a nonzero $R$. When the number of states is $N = 2$ and the index arithmetic is therefore (mod 2) the choice of sign does not in fact matter and we always get the 8-vertex ansatz. However, for $N > 2$ the two sign choices for $\epsilon\tau$ are genuinely different. The permutation matrix $R = \delta_i^l \delta_j^k$ is always included, but a diagonal $R$-matrix is included only with the above choice $\epsilon = -\tau = 1$. On the other hand, the choice $\epsilon = \tau = 1$ gives a manifestly left-right symmetric restriction.

## 4 The tetrahedron equation

We will now derive the tetrahedron equations for the three possible labeling schemes using the above derivation of star-triangle and Yang-Baxter equations as a guide.

### 4.1 The basic state and the basic scattering process

To begin with we have a two-dimensional ambient space and in it one-dimensional objects, straight strings. But this is not yet the basic state, for it we need two intersecting strings as shown in Fig. 5. The two strings divide the space into four quadrants and they also cut each other in two parts, furthermore we have a particle at the intersection of the strings.

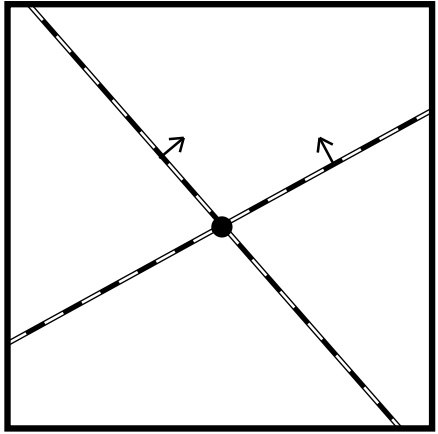

Figure 5: The basic state of two intersecting strings

Thus to fully label the basic state we need four vacuum labels, four string labels and one particle label (along with two dynamical characterizations, the string velocity vectors, which also define the motion of the particle).

The idea that the position of a particle is defined by intersecting strings has a precedent in soliton physics. The 2+1 dimensional Davey-Stewartson equation has localized solitons (called 'dromions') that are exactly of this type. It turns out that in this (and some other) soliton systems one can separate a physical and a 'ghost'-fields. The ghosts are normal plane-wave solitons but they do not show up in the physical field. However, at the intersections of the plane-wave ghosts the physical field is excited, creating an object that is localized in two dimensions [12].

For the basic scattering process we have to add one more straight string. The three strings form a triangle and the initial and final states of the basic scattering process look as in Fig. 6. First the triangle formed by the three strings decreases to a point, turns over and then starts to expand again. A perspective view of the process is given in Fig. 7. During scattering the string velocities do not change (elasticity) but the labelings of the inverting triangle can change, that is, the vacuum at the center, the three string segments at the sides, and the three particles at the corners.

## 4.2 The three labeling schemes

The scattering process of Figures 6 and 7 is characterized by a scattering amplitude which depends on the various labels given in Fig. 6. Some of the labels may of course be ignored. We will mainly discuss the following three labeling schemes:

1. If we are only interested in the state of the vacua the scattering amplitude is given by [7]
$$w(a|efg|bcd|h), \qquad (11)$$
   where the rule is to write first the vacuum in the center of the initial state (c.f. Fig. 6), then the vacua having a common edge with the center of the initial state

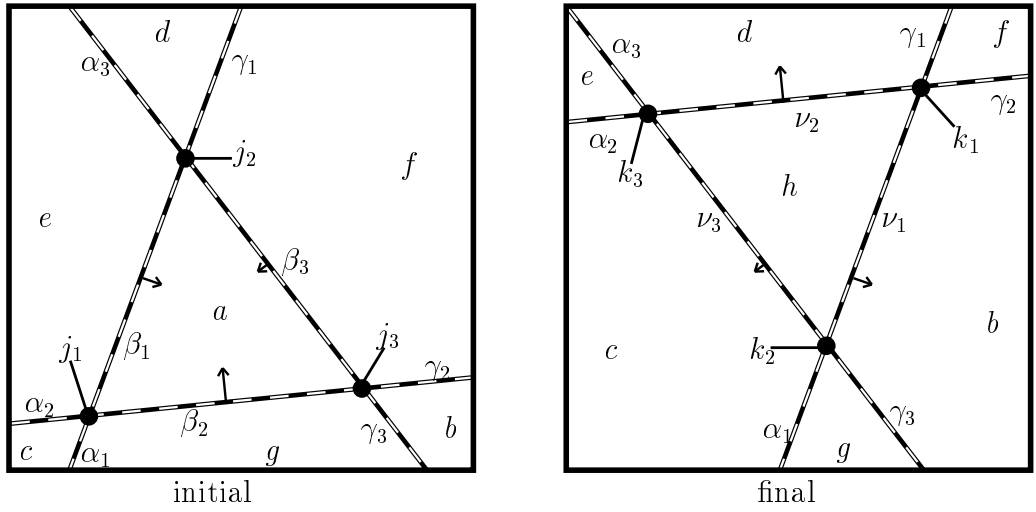

Figure 6: The initial and final states of the basic scattering process of three strings.

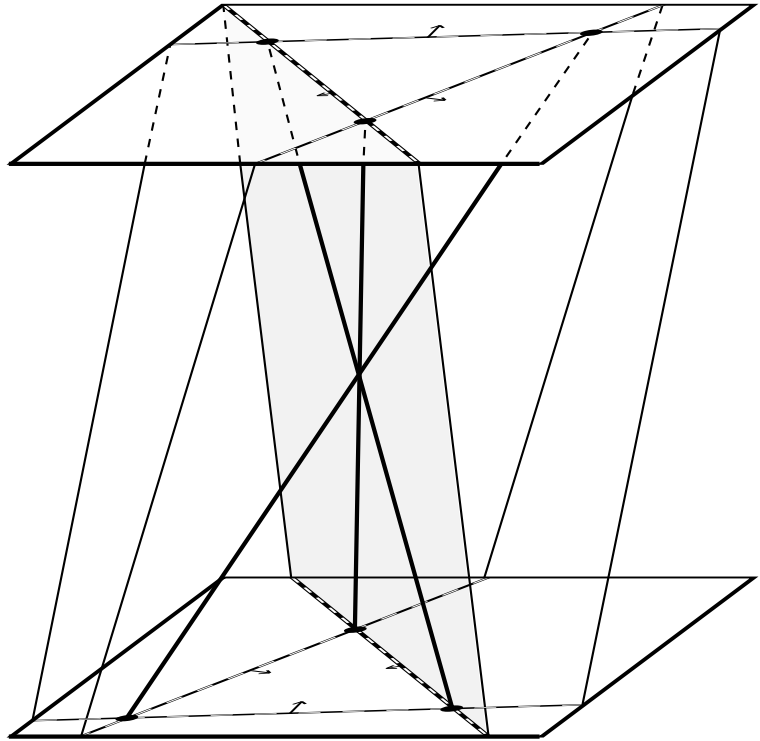

Figure 7: A perspective view of the scattering of three strings, the vertical axis is for time.

(clockwise, starting from the left), the vacua having a common edge with the center of the final state (clockwise, starting from right) and finally the new vacuum at the center of the final state. (Some authors use counterclockwise labeling [4].) For vacua we use lower case letters from the beginning of the alphabet. If we look at the system in three dimensions with time as the third axis then vacuum labeling can also be called volume or cell labeling.

2. In string (or face) labeling we write the process as

$$S\,{}^{\alpha_1\beta_1\gamma_1\nu_1}_{\alpha_2\beta_2\gamma_2\nu_2}_{\alpha_3\beta_3\gamma_3\nu_3}\,, \qquad (12)$$

where the rule is to write on each row the labels of one string, starting from the bottom of the leftmost string in the initial state and then clockwise (see Fig. 6). The last entry in each section is for the string segment in the middle of the new triangle in the final state. For strings we use greek letters. (The convention in [1] is slightly different)

3. In particle (or edge) labeling we write the scattering amplitude as [2]

$$R^{k_1 k_2 k_3}_{j_1 j_2 j_3}. \qquad (13)$$

Here the lower indices give the corners of the triangle in the initial state, (clockwise, starting from lower left-hand corner) and the upper indices refer to the triangle in the final state. For particles we use lower case letters from the middle of the alphabet.

## 4.3 The factorization condition

In analogue to Sec. 2.2 we get a factorization condition when we add one more element into the picture. With the fourth straight string the initial state can always be rotated into the one given in Fig. 8; the four strings make a arrowhead-like figure that points upwards. We can now go to a frame where the top of the arrow stays fixed and observe the motion of the intersection of the other two strings. Depending on the relative initial positions of the stings this intersection can pass the stationary intersection on the left or on the right. Triple intersections take place whenever either of these intersections crosses a third string. But on the right and left hand sides of Fig. 8 the order of the triple intersections is different; the tetrahedron equation is obtained when we require that both scenarios should give the same result.

A three dimensional space-time description of this is given in Fig. 9. At the center is *the* tetrahedron of the tetrahedron equation, its corners have been marked by black dots. The tetrahedron has different orientation on the different sides of the equation in the same way that the triangle is turned over in Fig. 4. In the following we are interested on the labeling associated with the scattering process. For that purpose it is easier to look at a sequence of two dimensional sections rather than the three-dimensional tetrahedron, the five essentially different sections are given in Figures 10 and 11.

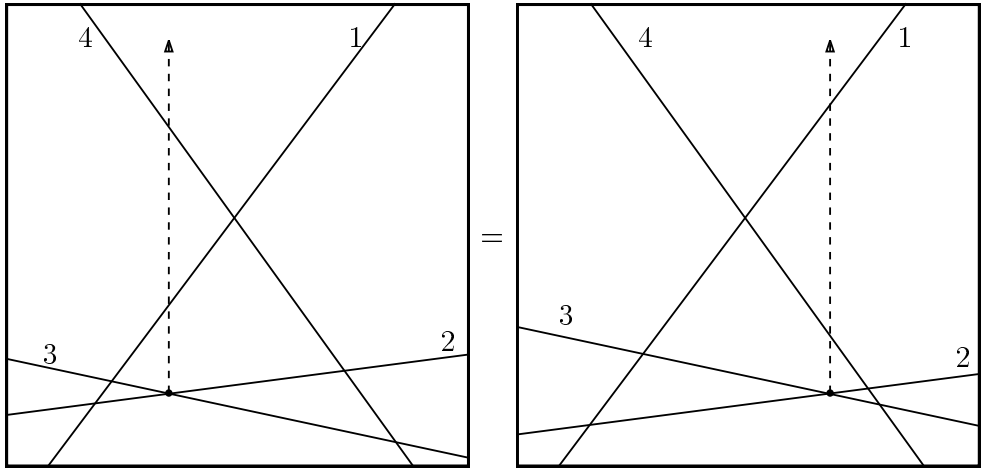

Figure 8: The two scattering scenarios with four strings. The frame of reference has been chosen so that the intersection of strings 1 and 4 is stationary. The intersection point of strings 2 and 3 can now pass the 1-4 intersection point on the left or on the right and both scenarios should give the same result.

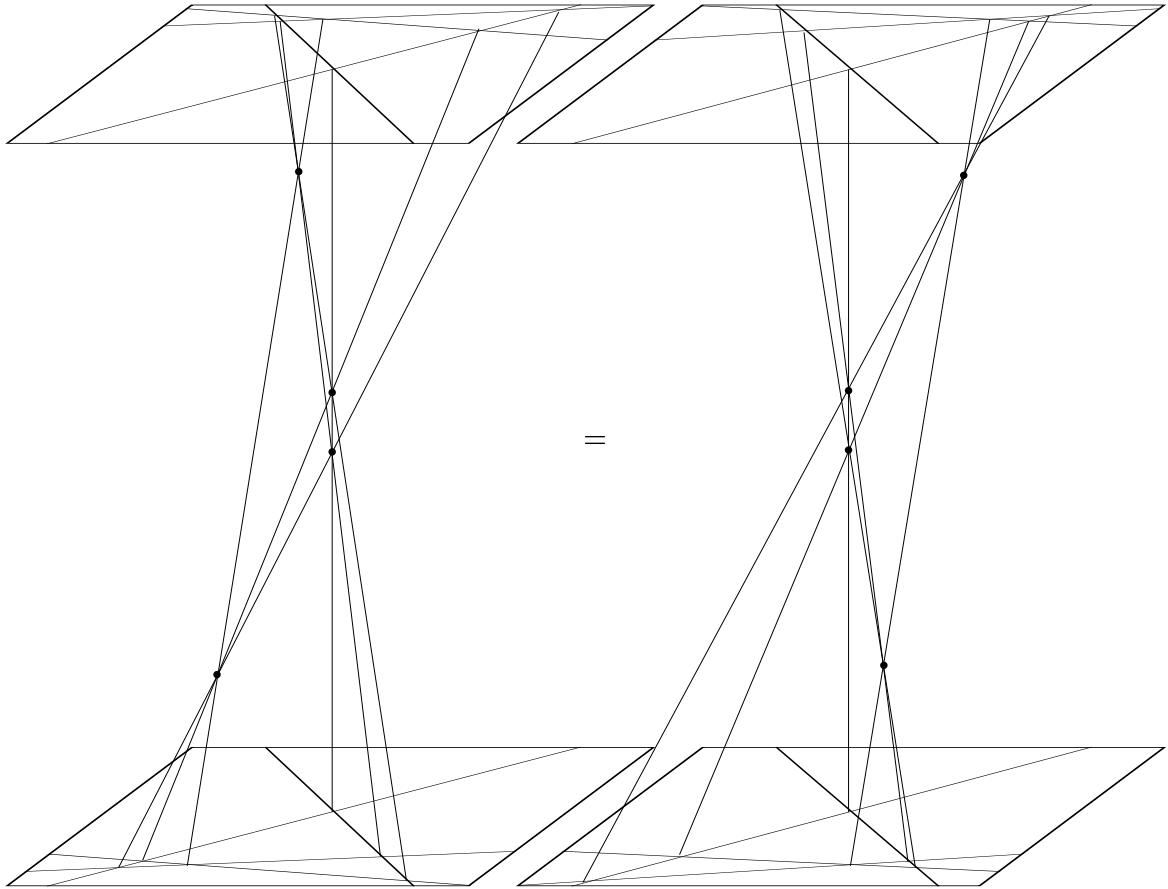

Figure 9: A three dimensional view of the two possible scattering scenarios

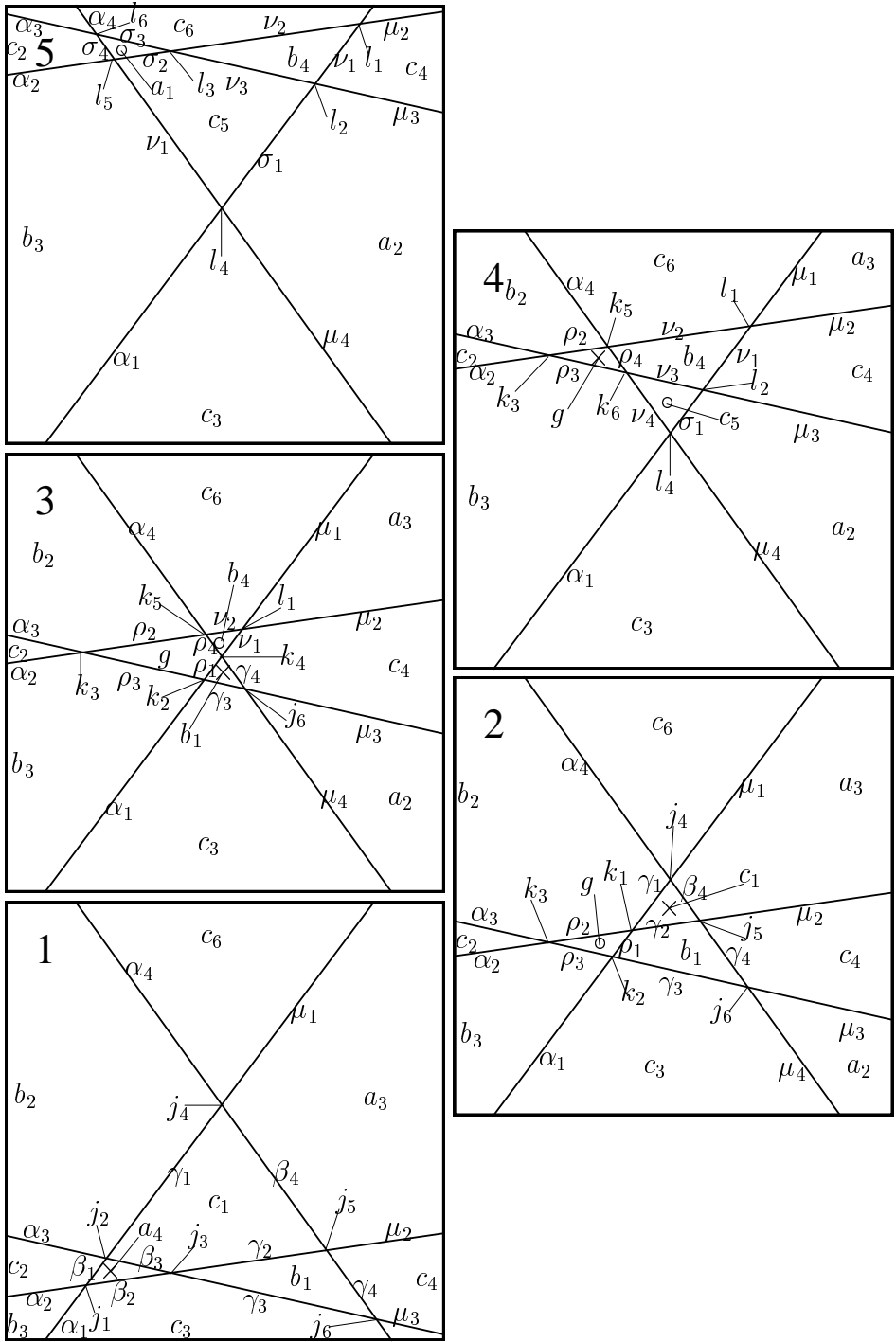

Figure 10: The sequence of five different time-slices of four string scattering corresponding to the LHS of Figure 9. The triangle that is about to turn over is marked with a cross, the result with a circle. All labels within and bordering the triangle change.

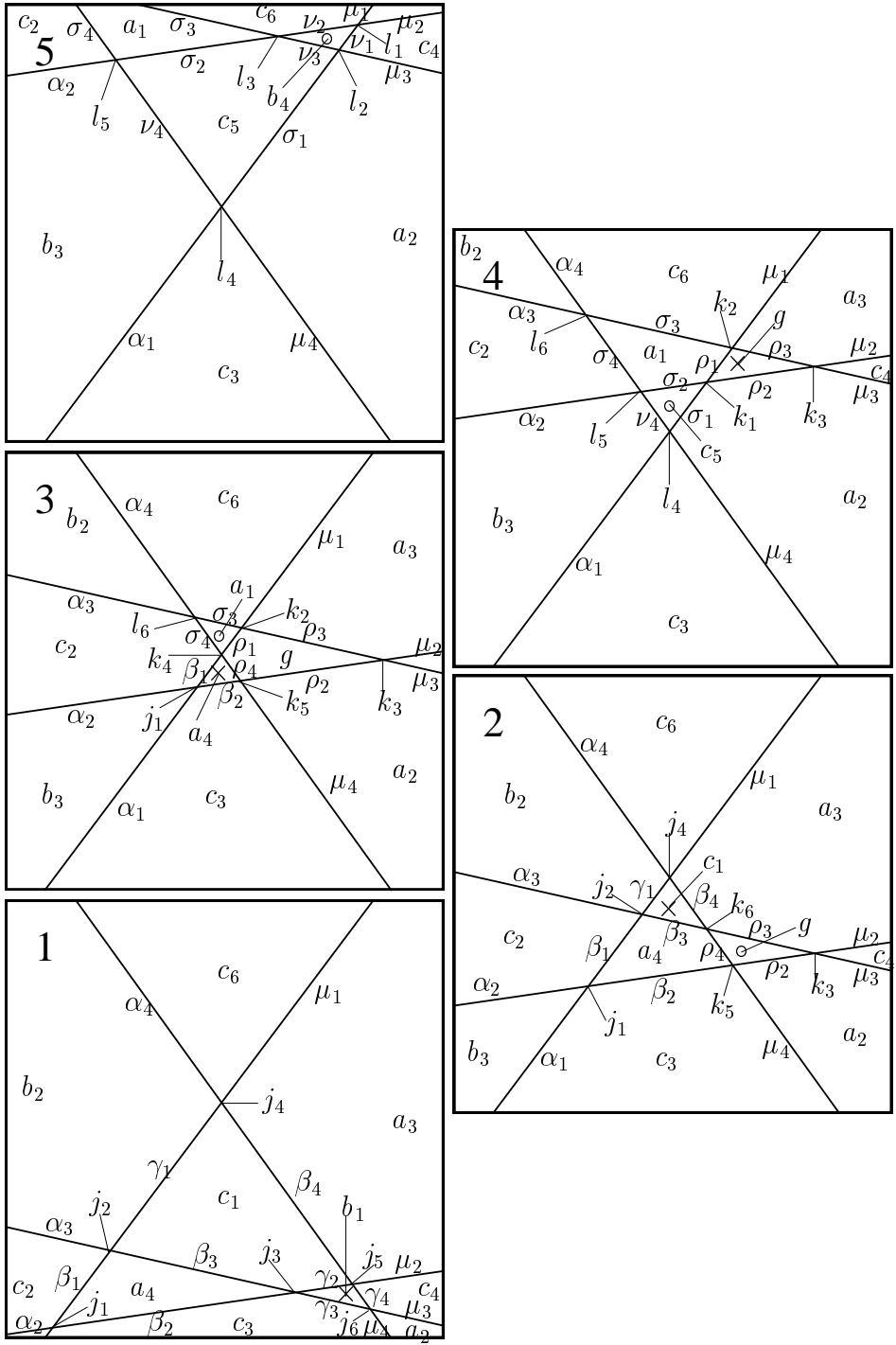

Figure 11: The sequence of time-slices corresponding to the RHS of Figure 9.

Let us first consider the vacuum or cell labeling. Then following the diagrams of Figure 10 and 11 using the convention of Eq. (11) we get the $d^{14}$ equations (Eq. (2.2) of [7])

$$\sum_g w(a_4|c_2c_1c_3|b_1b_3b_2|g)w(c_1|b_2a_3b_1|c_4gc_6|b_4)w(b_1|gc_4c_3|a_2b_3b_4|c_5)w(g|b_2b_4b_3|c_5c_2c_6|a_1)$$
$$=\tag{14}$$
$$\sum_g w(b_1|c_1c_4c_3|a_2a_4a_3|g)w(c_1|b_2a_3a_4|gc_2c_6|a_1)w(a_4|c_2gc_3|a_2b_3a_1|c_5)w(g|a_1a_3a_2|c_4c_5c_6|b_4).$$

If instead we use string labeling (12) we get $d^{24}$ equations

$$\sum_{\rho_1\rho_2\rho_3\rho_4} S^{\alpha_1\beta_1\gamma_1\rho_1}_{\alpha_2\beta_2\gamma_2\rho_2}{}_{\alpha_3\beta_3\gamma_3\rho_3} S^{\rho_1\gamma_1\mu_1\nu_1}_{\rho_2\gamma_2\mu_2\nu_2}{}_{\alpha_4\beta_4\gamma_4\rho_4} S^{\alpha_1\rho_1\nu_1\sigma_1}_{\rho_3\gamma_3\mu_3\nu_3}{}_{\rho_4\gamma_4\mu_4\nu_4} S^{\alpha_2\rho_2\nu_2\sigma_2}_{\alpha_3\rho_3\nu_3\sigma_3}{}_{\alpha_4\rho_4\nu_4\sigma_4}$$
$$= \sum_{\rho_1\rho_2\rho_3\rho_4} S^{\beta_2\gamma_2\mu_2\rho_2}_{\beta_3\gamma_3\mu_3\rho_3}{}_{\beta_4\gamma_4\mu_4\rho_4} S^{\beta_1\gamma_1\mu_1\rho_1}_{\alpha_3\beta_3\rho_3\sigma_3}{}_{\alpha_4\beta_4\rho_4\sigma_4} S^{\alpha_1\beta_1\rho_1\sigma_1}_{\alpha_2\beta_2\rho_2\sigma_2}{}_{\sigma_4\rho_4\mu_4\nu_4} S^{\sigma_1\rho_1\mu_1\nu_1}_{\sigma_2\rho_2\mu_2\nu_2}{}_{\sigma_3\rho_3\mu_3\nu_3} . \tag{15}$$

This is the original tetrahedron equation, Eq. (3.9) of [1], up to some label rearrangements.

Finally in the particle or edge labeling we get $d^{12}$ equations [2]

$$R^{k_1k_2k_3}_{j_1j_2j_3} R^{l_1k_4k_5}_{k_1j_4j_5} R^{l_2l_4k_6}_{k_2k_4j_6} R^{l_3l_5l_6}_{k_3k_5k_6} = R^{k_3k_5k_6}_{j_3j_5j_6} R^{k_2k_4l_6}_{j_2j_4k_6} R^{k_1l_4l_5}_{j_1k_4k_5} R^{l_1l_2l_3}_{k_1k_2k_3}, \tag{16}$$

where we have used the Einstein summation convention over the repeated $k$ indices.

At this point we would like to discuss briefly how the Frenkel-Moore equation [3] is related to the tetrahedron labelings. This set of $d^8$ equations is given by

$$F^{k_1k_2k_3}_{j_1j_2j_3} F^{l_1l_2k_4}_{k_1k_2j_4} F^{n_1l_3l_4}_{l_1k_3k_4} F^{n_2n_3n_4}_{l_2l_3l_4} = F^{k_2k_3k_4}_{j_2j_3j_4} F^{k_1l_3l_4}_{j_1k_3k_4} F^{l_1l_2n_4}_{k_1k_2l_4} F^{n_1n_2n_3}_{l_1l_2l_3}, \tag{17}$$

where summation is over the repeated $k$ and $l$ indices. Note that when the rows in the string labeling (15) are turned into columns the index *numbering* in (15) and (17) is identical. In fact, it turns out that the Frenkel-Moore equation is obtained with string labeling, but the labeling is *non-local*: Let us assign to each string a global index, which changes whenever the string takes part in a triple collision. When one now goes through the sequence of scatterings in Figures 10 and 11 with this convention one obtains exactly (17). Such global scheme would not work for Yang-Baxter equation, or for cell or particle labeling for tetrahedron equation, but for still higher dimensional simplex equations there could be several different non-local labeling schemes.

## 5 Dualities between the labeling schemes

We will now construct the tetrahedron analogues of the Wu-Kadanoff duality between the three labeling schemes. We follow closely the derivation used in Sec. 3.

The general setting is that the Boltzmann weights map as $w : I_v^{\otimes 8} \to \mathbb{C}$ while equation (14) itself is defined on $I_v^{\otimes 15}$, the string scattering amplitudes as $S : I_s^{\otimes 12} \to \mathbb{C}$ and (15) is defined on $I_v^{\otimes 26}$, and finally for the particle scattering amplitudes we have $R : I_p^{\otimes 6} \to \mathbb{C}$ and equation (16) is defined on $I_p^{\otimes 18}$. We must now construct maps $F$ that connect the pairs as follows: $w = R \circ F_{vp}$, $w = S \circ F_{vs}$, and $S = R \circ F_{sp}$.

## 5.1 Duality between vacuum and particle labelings

The first step is to express the scattering amplitude $w$ of (11) in terms of the $R$ of (13). According to the principles of hierarchy, locality and linearity each index of $R$ is given by a linear combination of the vacua (cells) around the corresponding particle (edge). From Figure 6 we get, after checking also that labels match [step i) in 3.2]

$$w(a|efg|b\,c\,d|h) = R^{\alpha d+\beta h+\gamma b+\delta f,\alpha h+\beta c+\gamma g+\delta b,\alpha d+\beta e+\gamma c+\delta h}_{\alpha e+\beta c+\gamma g+\delta a,\alpha d+\beta e+\gamma a+\delta f,\alpha f+\beta a+\gamma g+\delta b}, \qquad (18)$$

from which $F_{vp}$ can be read off. The constants $\alpha$, $\beta$, $\gamma$, $\delta$ are the parameters of the map $F_{vp}$ and we will later find restrictions on them.

As for step ii), constraints on $R$, we observe that in (16) there are 6 summation indices, and in (14) only one, thus we must get at least 5 conditions through the 4 $R$'s, that is, one condition is then not enough but we need two for each $R$. The constraints can only be consequences of the particular form of $R$ is (18). Depending on the choice of parameters $\alpha$, $\beta$, $\gamma$, $\delta$ one can obtain linear relations among the labels of $R$ and we can then define $R = 0$ when these relations do not hold:

$$R^{lmn}_{ijk} = 0, \text{ unless } A_1 i + A_2 j + A_3 k + B_1 l + B_2 m + B_3 n = 0.$$

Using (18) and requiring that these linear relations be true for all values of $a, \ldots, h$ yields 8 equations. These can be easily solved and the interesting results grouped into two different cases, $VP_A$: $\alpha\gamma = \beta\delta$ and $VP_B$: $\alpha = \gamma = 0$, $\beta + \delta = 0$, which we now discuss separately. [There are also some solutions with only one nonzero parameter, but such a dependence is probably too simple to lead into interesting relationships and we will not pursue them further.]

### 5.1.1 Case $VP_A$ with all parameters nonzero

In the first case we find that if $\alpha\gamma = \beta\delta$ then there are two relationships between the labels of $R$ and thus we can define

$$R^{lmn}_{ijk} = 0, \text{ unless } \alpha l + \beta i = \beta m + \alpha j \text{ and } \gamma m + \beta j = \beta n + \gamma k. \qquad (19)$$

To get the corresponding condition for $w$ we may assume that the $l$ and $n$ labels of $R$ in (19) are determined from the others, and then we get

$$w(a|efg|bcd|h) = \bar{w}_A(\alpha e+\beta c+\gamma g+\delta a, \alpha d+\beta e+\gamma a+\delta f, \alpha f+\beta a+\gamma g+\delta b, \alpha h+\beta c+\gamma g+\delta b), \qquad (20)$$

or

$$R^{*m*}_{ijk} = \bar{w}_A(i,j,k,m). \qquad (21)$$

To prove that a duality exists between (14) and (16) we substitute (18) into (14) and convert it to have particle labels of (16). (We also use $\alpha\gamma = \beta\delta$ to eliminate $\delta$.) In the process one finds restrictions on the labels, five of the eight conditions relate $k_i$, on the LHS as

$$\alpha k_1 + \beta j_1 = \alpha j_2 + \beta k_2, \ \gamma k_2 + \beta j_2 = \gamma j_3 + \beta k_3, \ \alpha l_1 + \beta k_1 = \alpha j_4 + \beta k_4,$$
$$\gamma k_4 + \beta j_4 = \gamma j_5 + \beta k_5, \ \gamma l_4 + \beta k_4 = \gamma j_6 + \beta k_6. \qquad (22)$$

and on the RHS as

$$\alpha l_1 + \beta k_1 = \alpha k_2 + \beta l_2, \ \gamma l_2 + \beta k_2 = \gamma k_3 + \beta l_3, \ \gamma l_4 + \beta k_4 = \gamma k_5 + \beta l_5,$$
$$\alpha k_1 + \beta j_1 = \alpha k_4 + \beta l_4, \ \gamma k_4 + \beta j_4 = \gamma k_6 + \beta l_6. \tag{23}$$

and the remaining three restrict the unsummed indices:

$$\begin{aligned}
\alpha^2(j_4 - l_1) + \alpha\beta(l_2 - j_2) + \beta^2(j_1 - l_4) &= 0, \\
\alpha\beta(j_4 - l_3) + \beta\gamma(j_3 - l_4) + \alpha\gamma(l_2 - j_5) + \beta^2(l_5 - j_2) &= 0, \\
\beta^2(j_4 - l_6) + \beta\gamma(l_5 - j_5) + \gamma^2(j_6 - l_4) &= 0.
\end{aligned} \tag{24}$$

This means that the range of $F_{vp}^*$ is 10 dimensional and thus its kernel should be 5 dimensional. Indeed one finds during the label conversion that it can be generated by $\{a_1, c_1, c_2, c_3, c_6\}$.

After this it is possible to write the common equation to which (14) and (16) reduce under duality, in terms of $\bar{w}_A$ defined in (20) or (21) it reads

$$\begin{aligned}
&\sum_k \bar{w}_A(j_1, j_2, j_3, k) \bar{w}_A(j_2 + \bar{\beta}(k - j_1), j_4, j_5, l_2' + \bar{\beta}k) \bar{w}_A(k, l_2' + \bar{\beta}k, j_6, l_4) \\
&\qquad \times \bar{w}_A(j_2 + \bar{\gamma}(k - j_3), j_4 + \bar{\gamma}(l_2' - j_5) + \bar{\beta}\bar{\gamma}k, l_2' + \bar{\gamma}(l_4 - j_6) + \bar{\beta}k, l_5' + \bar{\gamma}l_4) \\
&= \sum_k \bar{w}_A(j_3, j_5, j_6, k) \bar{w}_A(j_2, j_4, j_5 + \bar{\gamma}(k - j_6), l_5' + \bar{\gamma}k) \bar{w}_A(j_1, l_5' + \bar{\gamma}k, k, l_4) \\
&\qquad \times \bar{w}_A(l_5' + \bar{\gamma}k + \bar{\beta}(l_4 - j_1), j_4 + \bar{\beta}(l_5' - j_2) + \bar{\beta}\bar{\gamma}k, j_5 + \bar{\beta}(k - j_3), l_2' + \bar{\beta}l_4),
\end{aligned} \tag{25}$$

[where we have used the further definitions $\beta = \alpha\bar{\beta}$, $\gamma = \beta\bar{\gamma}$, $l_2' = l_2 - \bar{\beta}l_4$, $l_5' = l_5 + \bar{\gamma}l_4$, and eliminated $l_1$, $l_3$, $l_6$ using (24)].

In this equation we have resolved all the conditions from vacuum–particle duality. The function $\bar{w}_A$ depends on only 4 labels, and in the equation there are altogether 9 free labels. Thus we have $d^9$ equations for $d^4$ functions. It has the original symmetry that the LHS and RHS are exchanged if we exchange the first and third variables of $\bar{w}_A$ and the labels as $1 \leftrightarrow 6$, $2 \leftrightarrow 5$, and change the parameters $\bar{\beta} \leftrightarrow \bar{\gamma}$.

### 5.1.2 Case $VP_A$ with some zero parameters

Above we assumed that all the parameters $\alpha$, $\beta$, $\gamma$, $\delta$ were nonzero. The form of (19) is such that $\beta$ must be nonzero (if not we must also have $\alpha$ or $\gamma$ vanishing and then there is only on constraint). We could, however, assume that $\alpha = \delta = 0$, which yields

$$R_{ijk}^{lmn} = 0, \text{ unless } i = m, \quad \gamma m + \beta j = \beta n + \gamma k, \tag{26}$$

or $\gamma = \delta = 0$, which gives

$$R_{ijk}^{lmn} = 0, \text{ unless } j = n, \quad \alpha l + \beta i = \beta m + \alpha j. \tag{27}$$

For both of these $\delta = 0$, but if $\delta \neq 0$ we can eliminate $\beta$ from (19) and write the conditions as

$$\delta l + \gamma i = \gamma m + \delta j, \quad \delta m + \alpha j = \alpha n + \delta k.$$

Now we get two further possibilities, for $\beta = \alpha = 0$

$$R_{ijk}^{lmn} = 0, \text{ unless } m = k, \quad \delta l + \gamma i = \gamma m + \delta j. \tag{28}$$

and for $\beta = \gamma = 0$

$$R_{ijk}^{lmn} = 0, \text{ unless } l = j, \quad \delta m + \alpha j = \alpha n + \delta k. \tag{29}$$

These four possibilities are really various reflection of each other, so it is sufficient to discuss only one of them here. Let us assume that $\gamma = \delta = 0$. In deriving (22)-(24) it was assumed only that $\beta$ and $\alpha$ were nonzero so we can use these results, the constraint (27) is also such that we can solve $l$ and $n$ in terms of the others. Thus the final result is just (25) with $\bar{\gamma} = 0$, i.e.

$$\begin{aligned}
&\sum_k \bar{w}_0(j_1, j_2, j_3, k) \bar{w}_0(j_2 + \bar{\beta}(k - j_1), j_4, j_5, l_2' + \bar{\beta} k) \bar{w}_0(k, l_2' + \bar{\beta} k, j_6, l_4) \\
&\quad \times \bar{w}_0(j_2, j_4, l_2' + \bar{\beta} k, l_5') \\
&= \sum_k \bar{w}_0(j_3, j_5, j_6, k) \bar{w}_0(j_2, j_4, j_5, l_5') \bar{w}_0(j_1, l_5', k, l_4) \\
&\quad \times \bar{w}_0(l_5' + \bar{\beta}(l_4 - j_1), j_4 + \bar{\beta}(l_5' - j_2), j_5 + \bar{\beta}(k - j_3), l_2' + \bar{\beta} l_4).
\end{aligned} \tag{30}$$

The results simplifies considerably, and the most interesting feature here is that on the RHS the second term does *not* contain the summation index.

The other three possibilities give similar results.

### 5.1.3 Case $VP_B$

For this solution the parameters are restricted by $\alpha = \gamma = 0$ and $\beta = -\delta$ (for simplicity we assume $\beta = 1$, $\delta = -1$), and thus

$$w(a|efg|b\,c\,d|h) = R_{c-a,e-f,a-b}^{h-f,c-b,e-h}, \tag{31}$$

and we can impose the constraints

$$R_{ijk}^{lmn} = 0, \text{ unless } m = i + k, \ j = l + n. \tag{32}$$

In this case we solve for $m$ and $j$ in terms of the others and define $\bar{w}_B$ by

$$w(a|efg|bcd|h) = \bar{w}_B(c - a, a - b, h - f, e - h) \text{ or } R_{i*k}^{l*n} = \bar{w}_B(i, k, l, n). \tag{33}$$

As in the previous case we can next work out the constraints this implies on the indices of the tetrahedron equation, on the LHS we get

$$k_2 = j_1 + j_3, \ k_3 + k_1 = j_2, \ k_4 = j_5 + k_1, \ k_5 = j_4 - l_1, k_6 = j_5 + k_1 - l_2, \tag{34}$$

and on the RHS

$$k_5 = j_3 + j_6, \ k_6 + k_3 = j_5, \ k_2 = l - 1 + l_3, \ k_4 + k_3 = j_2 + j_5, \ k_1 + k_3 = l_2, \tag{35}$$

while for the unsummed indices we get

$$l_4 = j_1 + j_3 + j_6, \ l_5 + l_2 = j_2 + j_5, \ j_4 = l_1 + l_3 + l_6. \tag{36}$$

The common equation in terms of $\bar{w}_B$ reads ($l_2 = l'_2 + j_5$, $j_4 = j'_4 + l_1 + l_6$)

$$\sum_k \bar{w}_B(j_1, j_3, k, j_2 - k) \bar{w}_B(k, j_5, l_1, j'_4 + l_6)$$
$$\times \bar{w}_B(j_1 + j_3, j_6, j_5 + l'_2, k - l'_2) \bar{w}_B(j_2 - k, k - l'_2, j'_4, l_6)$$
$$= \sum_k \bar{w}_B(j_3, j_6, j_5 - k, k) \bar{w}_B(j_2, k, j'_4 + l_1, l_6)$$
$$\times \bar{w}_B(j_1, j_3 + j_6, l'_2 + k, j_2 - l'_2) \bar{w}_B(l'_2 + k, j_5 - k, l_1, j'_4). \tag{37}$$

This has again the symmetry exchange symmetry mentioned before, now the LHS and RHS are exchanged if we exchange the first and the second, and third and fourth variables of $\bar{w}_B$ and the labels as $1 \leftrightarrow 6$, $2 \leftrightarrow 5$, except that $l'_2 \leftrightarrow -l'_2$.

## 5.2 Vacuum-string duality

To get a duality between vacuum and string labelings we assume (hierarchy!) that each string label is determined by the two neighboring vacuum labels, from Figure 6 and label matching we get [c.f. [7], Eq. (2.4)]

$$w(a|efg|bcd|h) = S\begin{smallmatrix} \epsilon c + \tau g, \epsilon e + \tau a, \epsilon d + \tau f, \epsilon h + \tau b \\ \epsilon e + \tau c, \epsilon a + \tau g, \epsilon f + \tau b, \epsilon d + \tau h \\ \epsilon d + \tau e, \epsilon f + \tau a, \epsilon b + \tau g, \epsilon h + \tau c \end{smallmatrix}, \tag{38}$$

which defines $F_{vs}$. Since there are only two free parameters we assume that both of them are nonzero. Using the same method as before we find that (38) implies five relations between the indices of $S$:

$$S\begin{smallmatrix} \alpha_1 \beta_1 \gamma_1 \nu_1 \\ \alpha_2 \beta_2 \gamma_2 \nu_2 \\ \alpha_3 \beta_3 \gamma_3 \nu_3 \end{smallmatrix} = 0, \quad \text{unless} \tag{39}$$

$$\tau\alpha_1 + \epsilon\beta_1 = \epsilon\alpha_2 + \tau\beta_2, \ \tau\beta_2 + \epsilon\gamma_2 = \epsilon\beta_3 + \tau\gamma_3, \ \tau\nu_1 + \epsilon\gamma_1 = \epsilon\nu_2 + \tau\gamma_2,$$
$$\tau\alpha_2 + \epsilon\nu_2 = \epsilon\alpha_3 + \tau\nu_3, \ \epsilon^2(\alpha_3 - \gamma_1) + \epsilon\tau(\gamma_2 - \alpha_2) + \tau^2(\alpha_1 - \gamma_3) = 0.$$

This reduces the number of relevant labels from 12 to 7. The conditions seem to be much more than we need to restrict the summation in (15) to that of (14), but when one observes the way the summation indices are distributed in (15) the conditions appear more reasonable.

We can use (39) to determine $\beta_2$, $\gamma_1$, $\gamma_3$, $\nu_1$ and $\nu_3$ and then we find that $w$ depends only on seven variables, we can take e.g.

$$w(a|efg|bcd|h) = W(\epsilon c + \tau g, \epsilon e + \tau c, \epsilon d + \tau e, \epsilon e + \tau a, \epsilon f + \tau a, \epsilon f + \tau b, \epsilon d + \tau h). \tag{40}$$

or

$$S\begin{smallmatrix} \alpha_1 \beta_1 * * \\ \alpha_2 * \gamma_2 \nu_2 \\ \alpha_3 \beta_3 * * \end{smallmatrix} = W(\alpha_1, \alpha_2, \alpha_3, \beta_1, \beta_3, \gamma_2, \nu_2). \tag{41}$$

When one now substitutes (38) into (14) and converts the vacuum labels to string labels it turns out that in the resulting expression the labels $\alpha_1, \alpha_2, \alpha_3, \alpha_4, \beta_1, \beta_3, \beta_4, \gamma_2, \mu_2, \mu_3, \nu_2, \nu_3, \sigma_3$ and one of the $\rho$'s can be taken free, all the others are then determined by the following equations:

$$-\alpha_3\epsilon + \beta_1\tau - \beta_3\tau + \epsilon\gamma_1 = 0, \ \alpha_1\tau - \alpha_2\epsilon + \beta_1\epsilon - \beta_2\tau = 0, \ -\beta_2\tau + \beta_3\epsilon - \epsilon\gamma_2 + \gamma_3\tau = 0,$$
$$-\epsilon\mu_1 + \epsilon\nu_2 + \mu_2\tau - \nu_1\tau = 0, \ \alpha_4\epsilon + \beta_4\tau - \epsilon\mu_1 - \gamma_1\tau = 0, \ \beta_4\epsilon - \epsilon\mu_2 - \gamma_2\tau + \gamma_4\tau = 0,$$
$$-\epsilon\nu_1 + \epsilon\nu_3 + \mu_3\tau - \sigma_1\tau = 0, \ \epsilon\gamma_4 - \epsilon\mu_3 - \gamma_3\tau + \mu_4\tau = 0, \ -\epsilon\nu_2 + \epsilon\sigma_3 + \nu_3\tau - \sigma_2\tau = 0,$$
$$-\alpha_1\tau + \epsilon\nu_4 - \epsilon\sigma_1 + \mu_4\tau = 0, \ -\alpha_3\tau + \alpha_4\epsilon - \epsilon\sigma_3 + \sigma_4\tau = 0, \tag{42}$$

for the free indices, while the $\rho$'s are related on the RHS as

$$-\epsilon\rho_1 + \epsilon\sigma_2 + \rho_2\tau - \sigma_1\tau = 0, \quad -\epsilon\mu_2 + \epsilon\rho_3 + \mu_3\tau - \rho_2\tau = 0, \quad -\beta_3\tau + \beta_4\epsilon - \epsilon\rho_3 + \rho_4\tau = 0, \quad (43)$$

and on the LHS by

$$-\epsilon\gamma_1 + \epsilon\rho_2 + \gamma_2\tau - \rho_1\tau = 0, \quad \alpha_2\tau - \alpha_3\epsilon + \epsilon\rho_2 - \rho_3\tau = 0, \quad -\alpha_4\epsilon + \epsilon\nu_2 + \rho_2\tau - \rho_4\tau = 0. \quad (44)$$

These conditions mean that the $d^{14}$ vacuum equations and $d^{24}$ string equations collapse to 'only' $d^{13}$ equations. Since the range is 14 dimensional the kernel is one dimensional, it can be generated by $c_5$.

Finally we can write down the common equation in terms of the intermediate function $W$ of (40) or (41) as follows ($\bar{\tau} = \tau/\epsilon$):

$$\begin{aligned}
\sum_k & W(\alpha_1, \beta_1 + \bar{\tau}(\alpha_1 - \beta_2), \alpha_3, \beta_1, \beta_3, \gamma_2, k) \\
& \times W((k - \alpha_3)/\bar{\tau} + \beta_1 - \beta_3 + \gamma_2, k, \nu_2 + \bar{\tau}(\alpha_3 - \nu_3) + \bar{\tau}^2(\nu_4 - \beta_1) + \bar{\tau}^3(\beta_2 - \alpha_1), \\
& \quad \alpha_3 + \bar{\tau}(\beta_3 - \beta_1), \beta_4, \beta_4 + \nu_1 - \nu_3 + \bar{\tau}(\nu_4 - \beta_3) + \bar{\tau}^2(\beta_2 - \alpha_1), \nu_2) \\
& \times W(\alpha_1, (k - \alpha_3)/\bar{\tau} + \beta_1 + \bar{\tau}(\alpha_1 - \beta_2), k - \alpha_3 + \nu_3 + \bar{\tau}(\beta_1 - \nu_4) + \bar{\tau}^2(\alpha_1 - \beta_2), \\
& \quad (k - \alpha_3)/\bar{\tau} + \beta_1 - \beta_3 + \gamma_2, (\nu_1 - \nu_3)/\bar{\tau} - \beta_3 + \gamma_2 + \nu_4 + \bar{\tau}(\beta_2 - \alpha_1), \mu_3, \nu_3) \\
& \times W(\beta_1 + \bar{\tau}(\alpha_1 - \beta_2), \alpha_3, \nu_2 + \bar{\tau}(\alpha_3 - \nu_3) + \bar{\tau}^2(\nu_4 - \beta_1) + \bar{\tau}^3(\beta_2 - \alpha_1), k, \\
& \quad k - \alpha_3 + \nu_3 + \bar{\tau}(\beta_1 - \nu_4) + \bar{\tau}^2(\alpha_1 - \beta_2), \nu_3, \sigma_3) \\
= \sum_k & W(\beta_2, \beta_3, \beta_4, \gamma_2, (\nu_1 - \nu_3)/\bar{\tau} - \beta_3 + \gamma_2 + \nu_4 + \bar{\tau}(\beta_2 - \alpha_1), \mu_3, k) \\
& \times W(\beta_1, \alpha_3, \nu_2 + \bar{\tau}(\alpha_3 - \nu_3) + \bar{\tau}^2(\nu_4 - \beta_1) + \bar{\tau}^3(\beta_2 - \alpha_1), \alpha_3 + \bar{\tau}(\beta_3 - \beta_1), \beta_4, k, \sigma_3) \\
& \times W(\alpha_1, \beta_1 + \bar{\tau}(\alpha_1 - \beta_2), (\sigma_3 - \nu_2)/\bar{\tau} + \nu_3 + \bar{\tau}(\beta_1 - \nu_4) + \bar{\tau}^2(\alpha_1 - \beta_2), \beta_1, (k - \beta_4)/\bar{\tau} \\
& \quad + \beta_3, (k - \beta_4 - \nu_1 + \nu_3)/\bar{\tau} + \beta_3 + \mu_3 - \nu_4 + \bar{\tau}(\alpha_1 - \beta_2), (\sigma_3 - \nu_2)/\bar{\tau} + \nu_3) \\
& \times W((\nu_3 - \nu_1)/\bar{\tau} + \mu_3, (\sigma_3 - \nu_2)/\bar{\tau} + \nu_3, \sigma_3, (\sigma_3 - \nu_2)/\bar{\tau} - \beta_4 + k + \nu_3 + \bar{\tau}(\beta_3 - \nu_4) \\
& \quad + \bar{\tau}^2(\alpha_1 - \beta_2), k, \beta_4 + \nu_1 - \nu_3 + \bar{\tau}(\nu_4 - \beta_3) + \bar{\tau}^2(\beta_2 - \alpha_1), \nu_2). \quad (45)
\end{aligned}$$

## 5.3 String-Particle duality

We proceed as before: First, by hierarchy, locality, linearity we find that $S$ and $R$ are related by

$$S^{\alpha_1\beta_1\gamma_1\nu_1}_{\substack{\alpha_2\beta_2\gamma_2\nu_2\\\alpha_3\beta_3\gamma_3\nu_3}} = R^{\xi\gamma_1+\zeta\nu_2+\omega\nu_1+\eta\gamma_2,\xi\nu_1+\zeta\nu_3+\omega\alpha_1+\eta\gamma_3,\xi\nu_2+\zeta\alpha_3+\omega\alpha_2+\eta\nu_3}_{\xi\beta_1+\zeta\alpha_2+\omega\alpha_1+\eta\beta_2,\xi\gamma_1+\zeta\alpha_3+\omega\beta_1+\eta\beta_3,\xi\gamma_2+\zeta\beta_3+\omega\beta_2+\eta\gamma_3}, \quad (46)$$

where the 24 free constants have been reduced to 4 by label matching. This defines $F_{sp} : I_s^{12} \to I_p^6$.

Next we must find the label restriction that (46) allows; the results can be grouped into two cases, which bear interesting relation with the two cases of vacuum–particle duality.

### 5.3.1 Case $SP_A$

The first solution is obtained if $\eta = \zeta = 0$, then we can impose the condition

$$R^{lmn}_{ijk} = 0, \text{ unless } \xi l + \omega i = \omega m + \xi j, \quad (47)$$

another solution follows if we take instead $\omega = \xi = 0$, and then we get

$$R^{lmn}_{ijk} = 0, \text{ unless } \eta m + \zeta j = \zeta n + \eta k. \tag{48}$$

One immediately recognizes these as the two parts of the case $VP_A$ in (19): in the first case one identifies $\xi = \alpha$, $\omega = \beta$ and in the second case $\zeta = \beta$, $\eta = \gamma$. Thus the string–particle duality follows (with proper parameter choices) from vacuum–particle duality, because for the first one we need to imposed only "half" of the conditions of the second one.

One can now work out the conditions of either half as has been done before, but we will not do so here, because their combination as given in $VP_A$ is in practice probably the only important one.

### 5.3.2 Case $SP_B$

The other case is obtained if we have $\zeta = -\xi$, $\eta = -\omega$, then we have the restriction

$$R^{lmn}_{ijk} = 0, \text{ unless } \omega(m - i - k) = -\xi(j - l - n), \tag{49}$$

This is a combination of the two conditions of case $VP_B$ (32), thus again the existence of vacuum–particle duality implies the existence of a corresponding string particle duality. The detailed computations proceed as follows.

If we use (49) to determine $j$ in terms of the other labels, we find the corresponding condition for $S$ as

$$S^{\alpha_1\beta_1\gamma_1\nu_1}_{\substack{\alpha_2\beta_2\gamma_2\nu_2\\ \alpha_3\beta_3\gamma_3\nu_3}} = \tilde{S}(\xi\beta_1 - \xi\alpha_2 + \omega\alpha_1 - \omega\beta_2, \xi\gamma_2 - \xi\beta_3 + \omega\beta_2 - \omega\gamma_3, \xi\gamma_1 - \xi\nu_2 + \omega\nu_1 - \omega\gamma_2,$$
$$\xi\nu_1 - \xi\nu_3 + \omega\alpha_1 - \omega\gamma_3, \xi\nu_2 - \xi\alpha_3 + \omega\alpha_2 - \omega\nu_3). \tag{50}$$

or

$$R^{lmn}_{i*k} = \tilde{S}(i, k, l, m, n). \tag{51}$$

Next we substitute (46) into (15) and convert to particle labeling. In the process we find that on the LHS the summation indices are restricted by

$$\xi(j_2 - k_1 - k_3) + \omega(-j_1 - j_3 + k_2) = 0, \ \xi(j_4 - k_5 - l_1) + \omega(-j_5 - k_1 + k_4) = 0,$$
$$\xi(k_4 - k_6 - l_2) + \omega(-j_6 - k_2 + l_4) = 0, \tag{52}$$

and on the RHS by

$$\xi(j_5 - k_3 - k_6) + \omega(-j_3 - j_6 + k_5) = 0, \ \xi(-j_4 + k_2 + l_6) + \omega(j_2 - k_4 + k_6) = 0,$$
$$\xi(-k_1 + k_4 - l_5) + \omega(-j_1 - k_5 + l_4) = 0, \tag{53}$$

and on the free labels on both sides we get one condition

$$\xi^2(j_4 - l_1 - l_3 - l_6) + \xi\omega(-j_2 - j_5 + l_2 + l_5) + \omega^2(j_1 + j_3 + j_6 - l_4) = 0. \tag{54}$$

This means that the range of $F^*_{sp}$ is 14 dimensional, and hence the kernel 10 dimensional. Indeed we find that the kernel is generated by $\{\alpha_4, \beta_3, \beta_4, \gamma_4, \rho_4, \nu_3, \nu_4, \mu_2, \mu_3, \mu_4\}$.

Finally we can write the common equation in terms of $\tilde{S}$ using (54) to eliminate $j_4$:

$$\sum_{k_1,k_3,k_5} \tilde{S}(j_1, j_3, k_1, j_1 + j_3 + \tfrac{\xi}{\omega}(k_1 + k_3 - j_2), k_3)$$
$$\times \tilde{S}(k_1, j_5, l_1, k_1 + k_3 + k_6 + l_2 - j_2 + \tfrac{\omega}{\xi}(j_1 + j_3 + j_6 - l_4), l_3 + l_6 + \tfrac{\omega}{\xi}(k_3 + k_6 - l_5))$$
$$\times \tilde{S}(j_1 + j_3 + \tfrac{\xi}{\omega}(k_1 + k_3 - j_2), j_6, l_2, l_4, k_6) \tilde{S}(k_3, k_6, l_3, l_5, l_6)$$
$$= \sum_{k_2,k_3,k_6} \tilde{S}(j_3, j_6, k_3, j_6 + j_3 + \tfrac{\xi}{\omega}(k_6 + k_3 - j_5), k_6)$$
$$\times \tilde{S}(j_2, k_6, l_3 + l_1 + \tfrac{\omega}{\xi}(k_3 + k_1 - l_2), k_1 + k_3 + k_6 + l_5 - j_5 + \tfrac{\omega}{\xi}(j_1 + j_3 + j_6 - l_4), l_6)$$
$$\times \tilde{S}(j_1, j_6 + j_3 + \tfrac{\xi}{\omega}(k_6 + k_3 - j_5), k_1, l_4, l_5) \tilde{S}(k_1, k_3, l_1, l_2, l_3). \tag{55}$$

This again has the reflections symmetry of the original equation.

## 6 Simultaneous dualities between all pairs

Above we have analyzed the dualities between pairs of labeling schemes. In this section we will discuss the possibility of a having simultaneously dualities between two or even three pairs of labelings. It is natural to assume that such pairs of dualities would imply relations between the parameters used to define the maps $F$, and we will try to satisfy any further conditions primarily by imposing relation on these parameters.

Let us first consider simultaneous existence of vacuum–particle and vacuum–string duality. For both dualities the conditions for their existence can be written as a condition on how the Boltzmann weights actually depend on the vacuum labels, (20) or (33) vs. (40). From these we can see that it will be possible to satisfy the requirements for vacuum–string duality as a consequence of vacuum–particle duality, if the right hand side of (20) (or (33)) can be expressed in terms of the combinations $\epsilon c + \tau g$ etc. on the right hand side of (40). For case $VP_A$ (for which $\alpha\gamma = \beta\delta$) it turns out that this can indeed be done, provided that

$$(\alpha\tau - \beta\epsilon)(\beta\tau - \gamma\epsilon) = 0. \tag{56}$$

This is a compatibility condition on the parameters used in the definition of the dualities and as said before we will first implement them. If (56) holds and we have vacuum-particle duality, we can then impose the restriction $W = \bar{w}_A$ and get at the same time vacuum–string duality. For example, if we use the first factor of (56) we get

$$W(x_1, x_2, x_3, x_4, x_5, x_6, x_7) = \bar{w}_A((\beta\tau x_2 + \epsilon\gamma(x_4 - x_2) + \gamma\tau x_1)/\tau^2, (\beta x_3 + \gamma x_5)/\tau,$$
$$(\beta\tau x_5 + \epsilon\gamma(x_4 - x_2 - x_5 + x_6) + \gamma\tau x_1)/\tau^2,$$
$$(\beta\epsilon(x_7 - x_3) + \beta\tau x_2 + \epsilon\gamma(x_4 - x_2 - x_5 + x_6) + \gamma\tau x_1)/\tau^2), \tag{57}$$

and this also implies that (45) becomes (25) (after a change of labels). For case $VP_B$ it turns out that no extra conditions on parameters are needed: $\bar{w}_B$ of (33) is automatically a function of the variables in $W$ and we can impose

$$W(x_1, x_2, x_3, x_4, x_5, x_6, x_7) = \bar{w}_B((x_2 - x_4)\beta/\tau, (x_5 - x_6)\beta/\tau,$$
$$(\epsilon(x_7 - x_3) + \tau(x_4 - x_5))\beta/(\epsilon\tau), (x_3 - x_7)\beta/\tau), \tag{58}$$

and as a consequence (45) reduces to (37). Thus in either case a vacuum–string duality can be obtained, once the vacuum–particle duality has been established.

For vacuum–particle and string–particle dualities the condition can be best written on $R$, (19) or (32) vs. (47), (48) or (49). But as we have already observed in Sec. 5.3, the first duality implies the second one when the parameters are chosen properly.

Next let us consider simultaneous vacuum–string and string–particle dualities. Here we are interested in the composition $F'_{vp} = F_{sp} \circ F_{vs}$, which defines a map from vacuum to particle lables. It can always be made and it is then interesting to see how it is related to the already discussed dualities of $VP_A$ and $VP_B$. In the first case of $SP_{A1}$ we put $\eta = \zeta = 0$ and then it turns out that the resulting $F'_{vp}$ is the same as that of $VP_A$ if we identify

$$\eta = \zeta = 0: \quad \alpha = \epsilon\xi,\ \beta = \epsilon\omega,\ \gamma = \omega\tau,\ \delta = \tau\xi. \tag{59}$$

In particular we recover $\alpha\gamma = \beta\delta$, which is the defining relation of the map $VP_A$. For the other case $SP_{A2}$ we have to put $\omega = \xi = 0$ and then $VP_A$ again follows when we identify

$$\omega = \xi = 0: \quad \alpha = \epsilon\zeta,\ \beta = \tau\zeta,\ \gamma = \eta\tau,\ \delta = \epsilon\eta. \tag{60}$$

Recall that each of the two choices of $SP_A$ yielded only one of the two condition of $VP_A$, the extra restriction from simultaneous vacuum–string duality has now somehow created the missing one. Recall also that the simultaneous existence of vacuum–particle and vacuum–string dualities is possible for case $VP_A$ if condition (56) holds, it now turns out that the first factor of (56) vanishes automatically for (60), the second for (59). Finally, when the string–particle duality is given by $SP_B$ we first put $\eta = -\omega$ and $\zeta = -\xi$ and then the composed map is that of $VP_B$ (where we have $\alpha = \gamma = 0$), if

$$\eta + \omega = \zeta + \xi = \alpha = \gamma = 0: \quad \beta = (\epsilon\omega - \tau\xi), \delta = -(\epsilon\omega - \tau\xi). \tag{61}$$

Note in particular that in this case we get a nonvanishing condition on the parameters, we must assume that $\epsilon\omega - \tau\xi \neq 0$.

In summary, one can always find compatible parameter values so that all the dualities can coexist.

# 7  Discussion

In the first part of this paper we have derived, from the string scattering point of view, three kinds of tetrahedron equations corresponding vacuum (cell) labeling (14), string (face) labeling (15) and particle (edge) labeling (16).

In the second part we have studied the condition under which analogues of Wu-Kadanoff duality can be established between these labeling schemes. The final result is that there are two essentially different ways this can be done, and then with proper choice of the map parameters it is in fact possible to have a simultaneous duality between each of the three pairs of labelings. In terms of $R$ the conditions and resulting equation can be obtained already from vacuum–particle duality and for the two alternatives they are given in (19) and (32). The other dualities then follow if we choose the remaining parameters properly, i.e. according to (60-61). This is a very satisfactory result: the labeling schemes and dualities form a coherent structure where everything fits together well.

It now remains to see how the derived results really will help in finding interesting solution to the tetrahedron equations. Another direction for future work would be the extension of the present duality study to 4-simplex equations; this seems to be a formidable task due to the large number of labels needed and the difficulty in the visualization of the scattering process and the associated label changes.

# Acknowledgements

I would like to thank H.J. deVega and F. Nijhoff for discussions and for comments on the manuscript.

# References

[1] A.B. Zamolodchikov, Zh. Eksp. Teor. Fiz. **79**, 641 (1980) [Sov. Phys. JETP **52**, 325 (1980)].

[2] V.V. Bazhanov ans Yu.G. Stroganov, Teo. Mat. Fiz. **52**, 105 (1982) [Theor. Math. Phys. **52**, 685 (1982)].

[3] I. Frenkel and G. Moore, Commun. Math. Phys. **138**, 259 (1991).

[4] J.M. Maillet and F. Nijhoff, Phys. Lett. A **134**, 221 (1989); ibid *Multidimensional Integrable Lattices, Quantum Groups, and the D-Simplex Equations*, preprint CERN.TH-5595/89.

[5] J.S Carter and M. Saito, *On Formulation and Solutions of Simplex Equations*, preprint.

[6] Yu. I. Manin and V.V. Schechtman, in *Group Theoretical Methods in Physics*, vol. I, (Yurmala, 1985), p. 151. (VNU Sci. Press, Utrecht, 1986); R. Lawrence, in *Topological and Geometrical Methods in Field Theory*, J. Mickelsson and O. Pekonen (eds.), p. 429 (World Scientific, 1992); F.P. Michielsen and F.W. Nijhoff in *Quantum Topology*, L. Kauffman and R.S. Baadhio (eds.), p. 230 (World Scientific, 1993).

[7] R.J. Baxter, Commun. Math. Phys. **88**, 185 (1983).

[8] V.V. Bazhanov and R.J. Baxter, J. Stat. Phys. **69**, 453 (1992); R.M. Kashaev, V.V. Mangazeev and Yu.G. Stroganov, Int. J. Mod. Phys. A **8**, 1399 (1993); I.G. Korepanov, Commun. Math. Phys. **154**, 85 (1993); J. Hietarinta, J. Phys. A: Math. Gen. **26**, L9 (1993).

[9] J.B. McGuire, J. Math. Phys. **5**, 622 (1964); A.B. Zamolodchikov and Al.B. Zamolodchikov, Ann. Phys. **120**, 253 (1979).

[10] R.J. Baxter, *Exactly Solved Models in Statistical Mechanics*, (Academic Press, 1982)

[11] F.W. Wu, Phys. Rev. **B4**, 2312 (1971); L.P Kadanoff and F.J. Wegner, Phys. Rev. **B4**, 3989 (1971)

[12] A.S. Fokas and P.M. Santini, Phys. Rev. Lett **63**, 1329 (1989); J. Hietarinta, Phys. Lett. A **149**, 113 (1990).